\documentstyle[11pt,amssymb]{article}
\begin{document}

\newcommand{\iint}{{\int\hskip-3mm\int}}
\newcommand{\dddot}[1]{{\mathop{#1}\limits^{\vbox to 0pt{\kern 0pt
 \hbox{.{\kern-0.25mm}.{\kern-0.25mm}.}\vss}}}}
\renewcommand{\leq}{\leqslant}

\newcommand{\be}{\begin{equation}}
\newcommand{\ee}{\end{equation}}
\newcommand{\bes}{\begin{eqnarray}}
\newcommand{\ees}{\end{eqnarray}}
\newcommand{\eens}{\nonumber\end{eqnarray}}
\newcommand{\nl}{\nonumber\\}
\renewcommand{\/}{\over}
\renewcommand{\d}{\partial}

% Doublespace version
%\setlength{\jot}{7mm}
%\setlength{\abovedisplayskip}{7mm}
%\setlength{\belowdisplayskip}{7mm}

\newcommand{\nnl}{\nl[6mm]}
\newcommand{\enl}{\\[6mm]}
\newcommand{\nle}{\nl[-1.5mm]\\[-1.5mm]}
\newcommand{\nllb}[1]{\nl[-1.5mm]\label{#1}\\[-1.5mm]}
\newcommand{\bl}{&&\quad}
\newcommand{\ab}{\allowbreak}

\newcommand{\no}[1]{{\,:\kern-0.7mm #1\kern-1.2mm:\,}}

\newcommand{\half}{{1\/2}}
\newcommand{\PP}{{\cal P}}
\newcommand{\NN}{{\cal N}}
\newcommand{\CPP}{C^\infty(\PP)}
\newcommand{\ZZ}{{\Bbb Z}}
\newcommand{\adag}{a^\dagger}
\newcommand{\bdag}{b^\dagger}
\newcommand{\ups}{\upsilon}
\newcommand{\dlt}{\delta}
\newcommand{\la}{\lambda}
\newcommand{\rsum}{{r\in\ZZ+\half}}
\newcommand{\oj}{{\frak g}}

\topmargin 1.0cm

\newpage
\vspace*{-3cm}
\pagenumbering{arabic}
\begin{flushright}
{\tt math-ph/9905005}
\end{flushright}
\vspace{12mm}
\begin{center}
{\huge Constrained Fock spaces as Virasoro modules}\\[14mm]
\renewcommand{\baselinestretch}{1.2}
\renewcommand{\footnotesep}{10pt}
{\large T. A. Larsson\\
}
\vspace{12mm}
{\sl Vanadisv\"agen 29\\
S-113 23 Stockholm, Sweden}\\
email: tal@hdd.se
\end{center}
\vspace{3mm}
\begin{abstract}
The method of constrained Hamiltonian systems can be used to reduce 
Fock modules. It is applied to the Virasoro algebra, where a possibly
new realization is found.
\end{abstract}
\medskip
\renewcommand{\baselinestretch}{1.5}

\section{Constraints}
Consider the Virasoro algebra $Vir$,
\be
[L_m, L_n] \approx (n-m) L_{m+n} - {c\/12} (m^3-m)\dlt_{m+n},
\label{Vir}
\ee
where $m,n \in \ZZ$.
The purpose of this note is to show that some representations of $Vir$
can be found by applying the machinery of constrained Hamiltonian
systems \cite{Dir64,HT92}, suitably adapted to the representation
theory framework, to standard Fock modules. Some of the unconstrained
Fock modules may be new.

Let $\oj$ be a Lie algebra with elements $A,B,...$, and assume that there
is an embedding of $\oj$ into a graded Poisson algebra $\CPP$,
where $\PP$ is the phase space.
Let $P,R,..$ label constraints $\chi_P$, which may be
bosonic or fermionic; $(-)^P$ denotes the corresponding parity factor.
The equations $\chi_P\approx0$ define a surface in $\CPP$,
where weak equality (i.e. equality
modulo constraints) is denoted by $\approx$.
Constraints are second class if the Poisson bracket matrix 
$C_{PR} = [\chi_P,\chi_R]$ 
is invertible; otherwise, they are first class and generate a Lie algebra.
First class constraints are connected to gauge symmetries, 
and Hamiltonian systems with first class constraints are known as 
gauge systems \cite{HT92}.
Assume that all constraints are second class. Then the matrix 
$C_{PR}$ has an inverse, denoted by $\Delta^{PR}$.
Our sign convention is $(-)^R \Delta^{PR}C_{RS} = \dlt^P_S$.
The Dirac bracket
\be
[A,B]^* = [A,B] - (-)^R [A,\chi_P] \Delta^{PR} [\chi_R,B]
\label{Dirac}
\ee
defines a new graded Poisson bracket which is compatible with the 
constraints: $[A, \chi_R]^* = 0$ for every $A\in\oj$.
Of course, there is no guarantee that the operators $A,B$ still generate
the same Lie algebra under the Dirac brackets. A sufficient condition
for this is that the constraints are covariant in the sense that
$[A,\chi_P] = 0$ for every $A$. A less restrictive condition is often
possible. Usually, the constraints can be divided into two sets
$\chi_P = (\Phi_a, \Pi^a)$, such that 
$[\Phi^a, \Phi_b] \approx 0$.
The $\Phi^a$ are then first class, and  $\Pi_a$ are gauge conditions.
It is now sufficient that $[A, \Phi_a] \approx 0$, 
because the components of $\Delta^{PR}$ that involve $\Pi$'s on 
both sides vanish.

The factor space $\CPP/\NN$, where $\NN$ is the ideal generated by the
constraints, is the algebra of functions on the constraint surface,
with Poisson structure given by the Dirac bracket. Quantization amounts 
to replacing Dirac brackets by graded commutators and normal ordering.
In particular, if $\oj$ is the diffeomorphism algebra in one dimension,
we obtain reduced Fock representations of $Vir$.

\section{Scalar boson}
Consider a bosonic Virasoro primary field $a_m$ of zero conformal weight
and its canonical conjugate $\adag_n$ (of weight $1$), $m,n \in \ZZ$. 
\be
[\adag_m, a_n] = \dlt_{m+n}, \qquad
[\adag_m, \adag_n] = [a_m, a_n] = 0.
\ee
Set
\bes
B_m = \adag_m + Mma_m, &\qquad&
\chi_m = \adag_m - Mma_m, 
\label{ca}\\
K_m = \sum_{r=-\infty}^\infty r \no{ \adag_{m-r} a_r },  &\qquad&
L_m = K_m + \la(m+1)B_m,
\label{Lm}
\ees
where $M$ (interpretable as a mass) and $\la$ are parameters.
Then $L_m$ generate a Virasoro algebra with central charge 
$c=2-24M\la^2$. The following formulas are useful in the verification.
\bes
[\chi_m, \chi_n] &=& - [B_m, B_n] = 2Mm\dlt_{m+n}, 
\label{bb}\\
{[}B_m, \chi_n] &=& 0, \\
{[}K_m, a_n] &=& (m+n)a_{m+n}, \qquad 
[K_m, \adag_n] = n \adag_{m+n}, \\
{[}K_m, \chi_n] &=& n\chi_{m+n}, \qquad\qquad 
[K_m, B_n] = n B_{m+n}.
\ees
The modes $a_m$ can be eliminated by imposing the constraints 
$\chi_m \approx 0$. 
By virtue of (\ref{bb}), $\chi_0\approx 0$ is first class.
However, (\ref{ca}--\ref{Lm}) are independent of $a_0$, so we can
impose the extra constraint $\chi_{\bar0}=a_0\approx0$, 
making all constraints second class. 
The non-zero elements of the Poisson bracket matrix are
\bes
C_{mn} &=& [\chi_m,\chi_n] = 2Mm\dlt_{m+n}, \qquad (m\neq0) \nl
C_{0\bar0} &=& -C_{\bar00} =[\chi_0,a_0] = 1,
\ees
with inverse 
\bes
\Delta^{mn} &=& {-1\/2Mm}\dlt_{m+n}, \qquad (m\neq0) \nl
\Delta^{\bar00} &=& -\Delta^{0\bar0} = 1.
\ees
The Dirac brackets are
\be
[\adag_m,\adag_n]^* = 0 - \sum_{rs} (Mm\dlt_{m+r})({-1\/2Mr}\dlt_{r+s}) 
(Ms\dlt_{s+n}) = -{M\/2}m\dlt_{m+n},
\label{DBa}
\ee
when $m\neq0$. However, because
\be
[\adag_0,\adag_0]^* = [\adag_0,a_0]^* = [a_0,a_0]^* = 0,
\ee
(\ref{DBa}) holds for the zero mode as well. Now,
\bes
[L_m, \chi_n] &=& n \adag_{m+n} - Mn(m+n)a_{m+n} = n\chi_{m+n}
\approx 0, \nl
{[}L_m, a_0] &=& ma_m + \la \dlt_m \approx {1\/M}\adag_m + \la\dlt_m.
\ees
If we now solve the constraints $\chi_m \approx a_0 \approx 0$ for 
$\adag_m$, the Virasoro generators take the form
\be
L_m \approx \sum_{r=-\infty}^\infty {1\/M}\no{ \adag_{m-r}\adag_r } 
 + 2\la(m+1)\adag_m,
\label{Va}
\ee
where terms involving $\adag_0 = \chi_0 \approx 0$ should be skipped.
One verifies that (\ref{Va}) generates a Virasoro algebra with 
central charge $c=1-24M\la^2$ under the Dirac brackets (\ref{DBa});
this is the well-known Feigin-Fuks module 
(\cite{DF84,Tho84} and references therein).
Moreover, $\adag_n$ transforms as a one-dimensional Christoffel symbol
\be
[L_m,\adag_n]^* = n\adag_{m+n} - M\la m(m+1)\dlt_{m+n}.
\ee

\section{Scalar fermion}
Consider a fermionic Virasoro primary field $b_r$ of weight $\la$ 
and its canonical conjugate $\bdag_r$ (of weight $1-\la$). 
We have 
\bes
\{b_r,\bdag_s\} &=& \dlt_{r+s}, \qquad
\{b_r,b_s\} = \{\bdag_r,\bdag_s\} = 0, \\
L_m &=& -\sum_\rsum (-\la m+r) \no{ \bdag_{m-r} b_r }, \\
{[}L_m, b_r] &=& ((1-\la) m+r)b_{m+r}, \qquad 
[L_m, \bdag_r] = (\la m + r) \bdag_{m+r}, 
\ees
where anti-commutators are explicitly indicated. For simplicity, we
take $r,s\in \ZZ+1/2$, so there is no zero mode. 
The $L_m$ generate a Virasoro algebra with
central charge $c = -2(1-6\la+6\la^2)$.
If (and only if) $\la=1/2$ (i.e. $c=1$), we can eliminate the modes 
$\bdag_r$ by the constraints
\bes
\chi_r = b_r - \bdag_r  \approx 0,&& 
\label{cb}\\
C_{rs} = \{\chi_r, \chi_s\} = -2\dlt_{r+s}, &\qquad&
\Delta^{rs} = \half\dlt_{r+s}, \nl
\{\chi_r, b_s\} = -\dlt_{m+n}, &\qquad&
\{\chi_r, \bdag_s\} = \dlt_{m+n}.
\ees
The Dirac brackets are
\be
\{b_r, b_s\}^* = 0 + \sum_{kl} 
 (-\dlt_{r+k})(\half\dlt_{k+l})(-\dlt_{l+s})
 = \half\dlt_{r+s}
\label{DBb}
\ee
{F}rom
\be
[L_m, \chi_r] = ({m\/2}+r) \chi_{m+r} \approx 0,
\ee
it follows that 
\be
L_m = -\sum_\rsum (-{m\/2}+r) \no{ b_{m-r} b_r },
\label{Vb}
\ee
generate a Virasoro algebra with central charge $c=1/2$
under the Dirac bracket (\ref{DBb}).

\section{Conclusion}
It has been shown that the theory of constrained Hamiltonian 
systems can be applied to reduce Fock representations of the Virasoro
algebra. The explicit realizations (\ref{Va}) and (\ref{Vb}) are of
course well known, but the unconstrained companion of the Feigin-Fuks 
representation (\ref{Lm}) (with non-zero $\la$) may be new. 
The true value of the method is that it can be applied
to more complicated algebras, such as diffeomorphism and current
algebras in higher dimensions. In that case it can be much more natural to
describe a representation as a Fock module plus constraints, rather than
to explicitly solve the constraints \cite{Lar98}.

It appears that the examples reflect the spin-statistics theorem.
Namely, it follows from (\ref{DBa}) that (\ref{ca}) can only hold 
for bosons, and from (\ref{DBb}) that (\ref{cb}) can only hold for
fermions (otherwise $C_{rs}=0$). 
Thus, for bosons (fermions) the canonical momentum is 
proportional to the first (zeroth) time derivative of the field,
which is typical for integer (half-integer) spin.

\end{document}